\documentclass[aps,prd,nofootinbib,amsmath,amssymb,showpacs,superscriptaddress,column]{revtex4}
\usepackage{graphicx}
\usepackage{dcolumn}
\usepackage{bm}
\usepackage{amssymb}
\usepackage{latexsym}
\usepackage[colorlinks, linkcolor=blue, citecolor=blue, urlcolor=blue]{hyperref}

\newcommand{\be}{\begin{equation}}
\newcommand{\ee}{\end{equation}}

\def\bea{\begin{eqnarray}}
\def\eea{\end{eqnarray}}
\begin{document}

\title{Statefinder diagnosis for the extended holographic Ricci dark energy model without and with interaction}

\author{Fei Yu\footnote{Corresponding author}}
\email{yufei@sau.edu.cn} \affiliation{College of Sciences, Shenyang
Aerospace University, Shenyang 110136, China}
\author{Jing-Fei Zhang}
\affiliation{College of Sciences, Northeastern University, Shenyang
110004, China}

\begin{abstract}
We apply the statefinder diagnostic to the extended holographic
Ricci dark energy (ERDE) model without and with interaction to study
their behaviors. We plot the trajectories of various parameters for
different cases. It is shown that the non-interacting model does not
reach the LCDM point $\{1,0\}$ and the interacting one is favored,
because the interaction makes the evolution of the statefinder pair
$\{r,s\}$ quite different.
\end{abstract}



\maketitle

\section{Introduction}\label{sec:Intro}
A series of astronomical observations over the past decade indicate
that our universe is undergoing a state of accelerated
expansion~\cite{Riess:1998SN,Perlmutter:1999SN,Spergel:2003WMAP,Spergel:2007WMAP,Adelman:2008SDSS}.
The combined analysis of these observational data shows that the
present universe is dominated by an exotic component (about 73\%),
dubbed dark energy, which has negative pressure so as to accelerate
the expansion of the universe. And the rest 27\% is matter
components (cold dark matter plus the baryon) and negligible
radiation~\cite{Knop:2003SN,Komatsu:2009WMAP}. For the sake of
lending support to this abnormal phenomenon by theoretical argument,
physicists have done a great deal of work, including construction of
new dark energy
models~\cite{Sahni:2000ijmpd373,Peebles:2003rmp559,Copeland:2006ijmpd1753,Li:2011sd}
and modification of the gravity theory~\cite{Sotiriou:2008rmp}.
Thereinto the simplest one is the cosmological constant. Einstein
originally introduced this constant into his field equation for
getting a static cosmological solution in 1917, according to his
transcendental concept about the universe. But later on, it had been
unused for decades, of course, because the universe is not static.
Recently, along with the discovery of the cosmic accelerated
expansion by supernova data of 1998, the cosmological constant has
been put forth once again. It corresponds to the vacuum energy with
an equation of state $w=-1$. However, this model suffers from two
fundamental problems, which are the fine-tuning problem and the
cosmic coincidence problem~\cite{Weinberg:1989rmp1}. In addition,
theorists have brought forth various scalar field models of dark
energy~\cite{Copeland:2006ijmpd1753}, in which the equation of state
parameter $w$ is dependent on time, such as
quintessence~\cite{Zlatev:1999prl896},
phantom~\cite{Caldwell:2002plb23}, quintom~\cite{Feng:2005plb35},
tachyon~\cite{Padmanabhan:2002prd021301}, ghost
condensate~\cite{Piazza:2004jcap07004} and so on.

In essence the dark energy problem should be an issue of quantum
gravity, although there is no mature theory of quantum gravity with
so many things unknown and uncertain at present. Currently, making
an approach to dark energy in the frame of quantum gravity is the
famed holographic dark energy inspired by the holographic
principle~\cite{Hooft:9310026,Cohen:1999prl4971}, which leads to the
energy density $\rho_{de}=3c^2M_p^2L^{-2}$, where $c$ is an
introduced numerical constant, $M_p$ is the reduced Planck mass with
$M_p^2=(8\pi G)^{-1}$ and $L$ is the infrared (IR) cut-off. Among
series of the holographic models, it is proved that those with the
Hubble scale and the particle horizon as the IR cut-off cannot give
rise to the cosmic acceleration~\cite{Hsu:2004plb13,Li:2004plb1}.
Instead, Li takes the future event horizon as the IR cut-off,
leading to a successful holographic model~\cite{Li:2004plb1}.
However, since the application of the future event horizon means
that the history of dark energy depends on the future evolution of
the scale factor $a(t)$~\cite{Cai:2007plb228}, some other versions
of the holographic dark energy have been put forth, such as the
agegraphic dark energy model~\cite{Wei:2007ty} and the holographic
Ricci dark energy (RDE) model~\cite{Gao:2009prd043511}. Recently,
the RDE model has been extended to the following
form~\cite{Granda:2008plb275}
\begin{equation}\label{ghde}
\rho_{de}=3M_p^2(\alpha H^2+\beta\dot{H}),
\end{equation}
where $\alpha$ and $\beta$ are constants to be determined.
Obviously, this extended model can reduce to the
RDE~\cite{Gao:2009prd043511} for the case of $\alpha=2\beta$. In
this paper, we will focus on this extended Ricci dark energy (ERDE)
model.

So far dark energy models have grown in number. An approach is
claimed to differentiate between them. As is well known, the
statefinder is a sensitive and robust geometrical diagnostic of dark
energy, which is constructed using both the second and third
derivatives of the scale factor
$a(t)$~\cite{Sahni:2003jetpl201,Alam:2003mnras1057}. Let us consider
the general form for the scale factor of the universe
\begin{equation}\label{scaleexpansion}
a(t)=a(t_0)+\dot{a}|_0(t-t_0)+\frac{\ddot{a}|_0}{2}(t-t_0)^2
+\frac{\dddot{a}|_0}{6}(t-t_0)^3+\cdots.
\end{equation}
Since the cosmic accelerated expansion is a fairly recent
phenomenon, we can confine our attention to small values of
$|t-t_0|$ in Eq. (\ref{scaleexpansion}). Thus, define the
statefinder pair $\{r,s\}$ as
\begin{equation}
r \equiv \frac{\dddot{a}}{aH^3}, \ \ \ s \equiv
\frac{r-1}{3(q-1/2)}.
\end{equation}
We know that the Hubble parameter $H=\dot{a}/a$ is constructed using
the first derivatives of $a(t)$, while the deceleration parameter
$q=-\ddot{a}/(aH^2)$ the second. So $r$ is naturally the next step
beyond $H$ and $q$, and $s$ is a linear combination of $r$ and $q$.
At the present time, the statefinder diagnostic has already been
used to diagnose and discriminate behaviors of many dark energy
models~\cite{Zimdahl:2004grg1483,Zhang:2005plb1,Zhang:2005kj,Wu:2005ijmpd1873,Zhang:2005xk,Zhang:2004gc,
Wei:2007plb1,Chang:2007jcap01016,Chang:2008mpla269,Shao:2008mpla65,Huang:2008ass175,
Panotopoulos:2008npb66,Shojai:2009el30002,Tong:2009prd023503,Zhang:2010ijmpd21},
especially of the holographic dark energy
models~\cite{Zhang:2005ijmpd1597,Setare:2007jcap03007,Zhang:2008plb26,Feng:2008plb231}.

In this paper, we apply the statefinder diagnostic to the ERDE model
without~\cite{Granda:2008plb275} and with~\cite{Yu:2010plb263}
interaction and finally we will see the advantages of the
interacting model. In the next section we briefly review the
interacting ERDE model. In Section III we do diagnosis with the
statefinder to the ERDE model without and with interaction. The last
section is for conclusion.

\section{Brief review of the interacting ERDE model}\label{sec:model}
To begin with, we shall briefly review the interacting ERDE
model~\cite{Yu:2010plb263}. The conservation equations of energy
densities in the spatially homogeneous and isotropic universe read
\begin{eqnarray}
\dot{\rho}_{de}+3H(1+w)\rho_{de} &=& -Q, \label{deconservation}\\
\dot{\rho}_m+3H\rho_m &=& Q,
\end{eqnarray}
where $w$ is the equation of state parameter (EOS) of dark energy,
$Q$ denotes the interaction between dark energy and matter by the
form $Q=3bH(\rho_{de}+\rho_m)$ with $b$ the coupling
constant~\cite{Li:2009jcap12014}. Positive $b$ indicates that dark
energy decays to matter, whereas matter to dark energy for negative
$b$. When introducing the parameter $r_\rho=\rho_m/\rho_{de}$ as the
density ratio of matter to dark energy, $Q$ can be rewritten in the
form $Q=3b(1+r_\rho)H\rho_{de}$. Making use of the conservation
equations we can get
\begin{equation}\label{dotratio}
\dot{r}_\rho=3H\left[wr_\rho+b(1+r_\rho)^2\right].
\end{equation}

Moreover, the Friedmann equation is
\begin{equation}\label{Friedmann}
3M_p^2 H^2=\rho_{de}+\rho_m,
\end{equation}
and the derivative of $H$ with respect to time can be given,
\begin{equation}\label{dotH}
\dot{H}=-\frac{3}{2}H^2\left(1+\frac{w}{1+r_\rho}\right).
\end{equation}
Defining the fractional energy densities
$\Omega_{de}\equiv\rho_{de}/(3M_p^2 H^2)$ and
$\Omega_m\equiv\rho_m/(3M_p^2 H^2)$, the Friedmann equation reads
$\Omega_{de}+\Omega_m=1$. So $r_\rho$ also has the form
$r_\rho=\rho_m/\rho_{de}=\Omega_m/\Omega_{de}$, leading to
\begin{equation}
\Omega_{de}=\frac{1}{1+r_\rho}.
\end{equation}
Substituting Eqs. (\ref{ghde}) and (\ref{dotH}) into Eq.
(\ref{Friedmann}), we get the relationship between $w$ and $r_\rho$,
\begin{equation}\label{w}
w=\left(\frac{2\alpha}{3\beta}-1\right)(1+r_{\rho})-\frac{2}{3\beta}.
\end{equation}
Further, replacing $w$ in Eq. (\ref{dotratio}) by Eq. (\ref{w}), we
obtain a differential equation of $r_\rho$ with respect to $x=\ln
a$, i.e., $dr_\rho/f(r_\rho)=dx$ where
\begin{eqnarray}
f(r_\rho) &=& \left(\frac{2\alpha}{\beta}-3+3b\right)r_\rho^2+\left(\frac{2\alpha}{\beta}-3-\frac{2}{\beta}+6b\right)r_\rho+3b, \nonumber \\
&=& Cr_\rho^2+Br_\rho+A.
\end{eqnarray}
After integration, $r_\rho$ is gained in the form
\begin{equation}\label{ratio}
r_{\rho}(x)=\frac{\sqrt{-\Delta}\tanh\left[-\frac{\sqrt{-\Delta}}{2}(x+D)\right]-B}{2C},
\end{equation}
where $\Delta=4AC-B^2<0$ is the discriminant of the quadratic
polynomial $f(r_\rho)$ and the integration constant $D$ is
\begin{equation}
D=-\frac{2}{\sqrt{-\Delta}}\tanh^{-1}\left[\frac{2Cr_{\rho
0}+B}{\sqrt{-\Delta}}\right].
\end{equation}
In the process of integration, boundary conditions $w_0=-1$ and
$r_{\rho0}=\Omega_{m0}/\Omega_{de0}=0.27/0.73$ have been used and
the subscript ``0'' denotes the current values of the physical
quantities. They are well consistent with current
observations~\cite{Spergel:2007WMAP,Komatsu:10014538}. Under these
conditions $\alpha$ is determined by $\beta$,
\begin{equation}
\alpha=\frac{2+3\beta r_{\rho0}}{2(1+r_{\rho0})}.
\end{equation}
At this rate, through Eq. (\ref{ratio}), Eq. (\ref{w}) describes the
evolution of dark energy.

\section{Statefinder diagnosis for the ERDE model without and with interaction}\label{sec:stafd}

\begin{figure*}
\centering
\includegraphics[width=2.1 in]{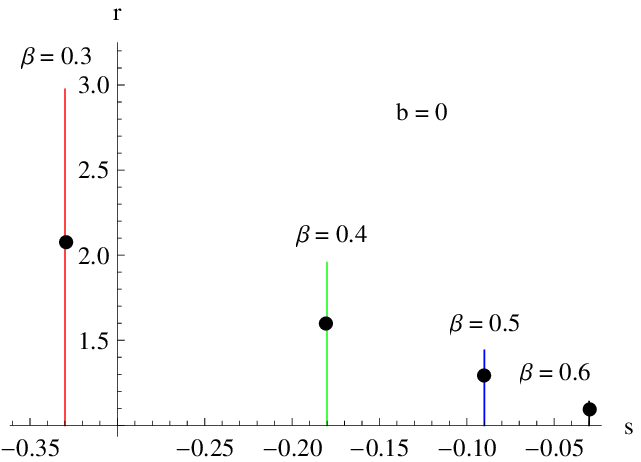}
\includegraphics[width=2.1 in]{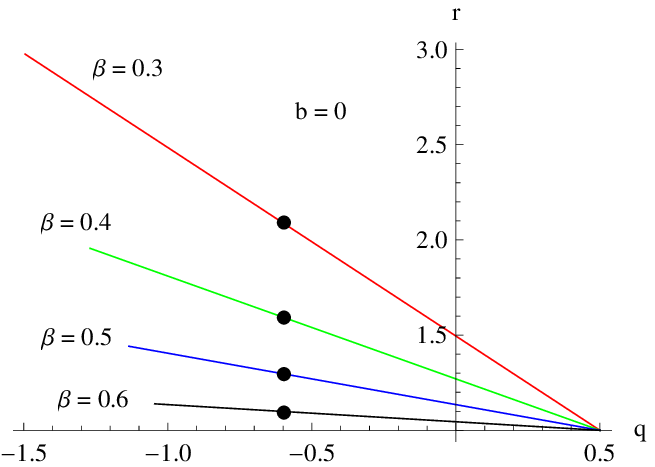}
\includegraphics[width=2.1 in]{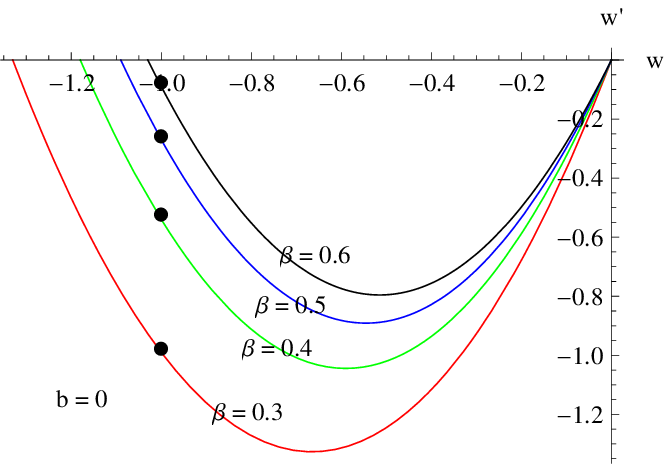}
\caption{(Color online.) The curves of $r(s)$, $r(q)$ and $w'(w)$ in
the ERDE model without interaction respectively in the $r-s$, $r-q$
and $w'-w$ planes for variable $\beta$. The dots denote today's
values of these parameters and $q_0=-0.595$ and $w_0=-1$ for all the
cases.}\label{nin}
\end{figure*}

\begin{figure*}
\centering
\includegraphics[width=2.1 in]{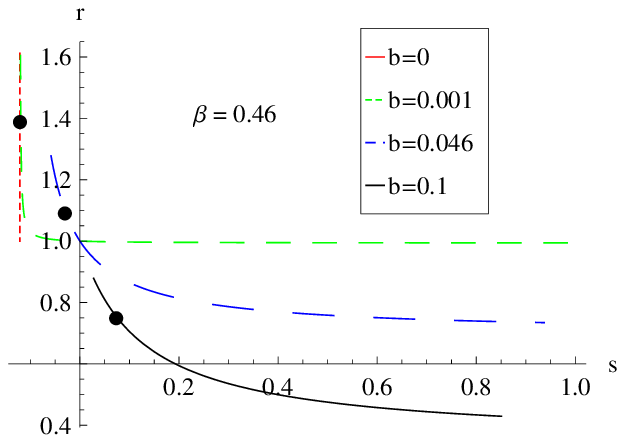}
\includegraphics[width=2.1 in]{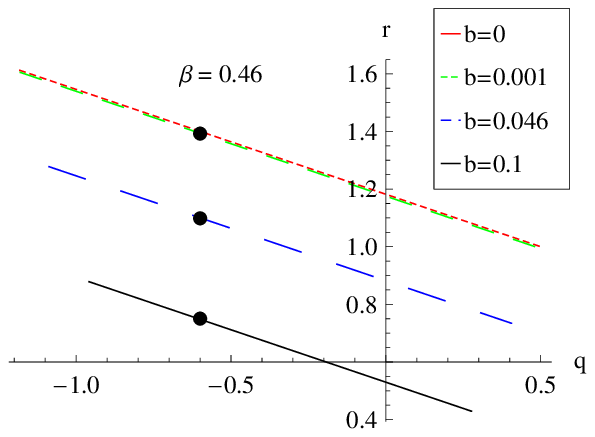}
\includegraphics[width=2.1 in]{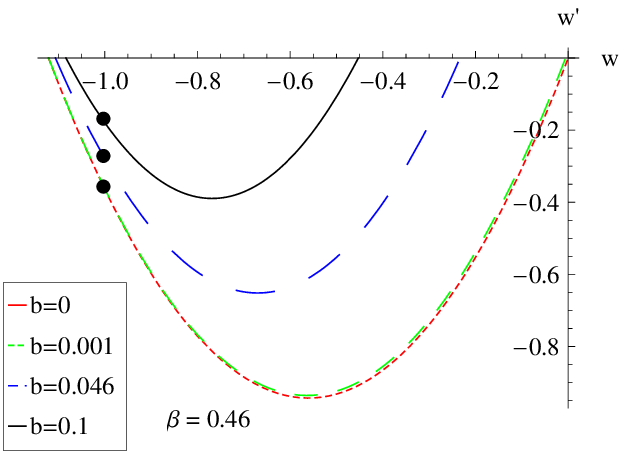}
\caption{(Color online.) The curves of $r(s)$, $r(q)$ and $w'(w)$ in
the ERDE model with interaction respectively in the $r-s$, $r-q$ and
$w'-w$ planes for variable $b$ with the best-fit $\beta=0.46$. The
dots denote today's values of these parameters and $q_0=-0.595$ and
$w_0=-1$ for all the cases.}\label{bf}
\end{figure*}

In what follows, we do diagnosis with the statefinder to the above
ERDE model. According to one of the basic dynamical equations of
cosmology
\begin{equation}
\frac{\ddot{a}}{a}=-\frac{4\pi G}{3}(\rho+3p),
\end{equation}
where $\rho$ and $p$ denote respectively the total energy density
and pressure of the universe, the statefinder parameters have the
form in terms of $\rho$ and $p$,
\begin{equation}
r=1+\frac{9(\rho+p)}{2\rho}\frac{\dot{p}}{\dot{\rho}}, \ \ \
s=\frac{(\rho+p)}{p}\frac{\dot{p}}{\dot{\rho}},
\end{equation}
as well as the deceleration parameter
\begin{equation}
q=-\frac{\ddot{a}}{aH^2}=\frac{1}{2}+\frac{3p}{2\rho}.
\end{equation}
Further, in view of $\rho=\rho_m+\rho_{de}$ and
$p=p_m+p_{de}=p_{de}=w\rho_{de}$, $\rho$ keeps conserved and
satisfies $\dot{\rho}=-3H(\rho+p)$ while
$\dot{p}=\dot{w}\rho_{de}+w\dot{\rho}_{de}$. Note that the
conservation equation of dark energy (\ref{deconservation}) is a
little more complicated, so we introduce the effective equation of
state of dark energy by
\begin{equation}
w^\textrm{eff}=w+b(1+r_\rho),
\end{equation}
then Eq. (\ref{deconservation}) recovers the standard form
\begin{equation}
\dot{\rho}_{de}+3H(1+w^\textrm{eff})\rho_{de}=0.
\end{equation}
So the statefinder and deceleration parameters can be expressed as
\begin{eqnarray}
r &=& 1-\frac{3}{2}\Omega_{de}\left[w'-3w(1+w^\textrm{eff})\right], \\
s &=& 1+w^\textrm{eff}-\frac{w'}{3w}, \\
q &=& \frac{1}{2}+\frac{3}{2}w\Omega_{de},
\end{eqnarray}
where `` $'$ '' denotes the derivative with respect to $x=\ln a$ and
$H=dx/dt$. When there is no interaction, i.e., $b=0$, we have
$w^{\rm eff}=w$. Therefore, the LCDM model with $w=-1$ leads to
constant statefinder parameters below
\begin{equation}
\{r,s\}|_{\rm LCDM}=\{1,0\}.
\end{equation}
This means that the LCDM model corresponds to a fixed point
$(s=0,r=1)$ in the statefinder $r-s$ plane. Thus, in virtue of this
feature, other models of dark energy can be measured for the
distance between them and the LCDM point to study their behaviors.

First, we plot curves of $r(s)$, $r(q)$ and $w'(w)$ in the ERDE
model without interaction respectively in the $r-s$, $r-q$ and
$w'-w$ planes in Fig.~\ref{nin}. In the $r-s$ plane, we see clearly
that $r$ is time dependent while $s$ constant. This is the same as
that in quiessence models (QCDM) in which $w={\rm constant}\neq -1$.
But what's different is shown in the $w'-w$ plane, i.e., $w$ is not
a constant. Especially when $\alpha=2\beta$, the model reduces to
RDE. The trajectories of statefinder parameters coincide with that
in Ref.~\cite{Feng:2008plb231}, namely, $\beta=0.5$ makes the
statefinder pair $\{r,s\}$ a fixed point $\{1,0\}$ corresponding to
LCDM model and the universe ends in a de Sitter phase corresponding
to the point $(w=-1,w'=0)$ in the $w'-w$ plane, also if $\beta<0.5$,
the trajectories will lie in the region $r>1$, $s<0$ while
$\beta>0.5$ in the region $r<1$, $s>0$. This is sort of different
from the general case (\ref{ghde}), in which $r$ starts from $1$ to
increase and $s<0$ for the fitting $\beta$, even for $\beta>0.5$.
Therefore it is concluded that in the non-interacting case the model
does not reach the LCDM point $\{1,0\}$.

\begin{figure*}
\centering
\includegraphics[width=2.1 in]{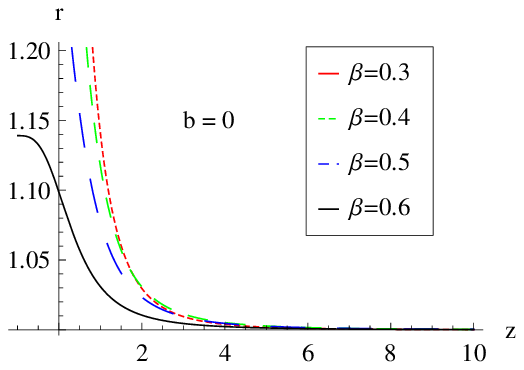}
\includegraphics[width=2.1 in]{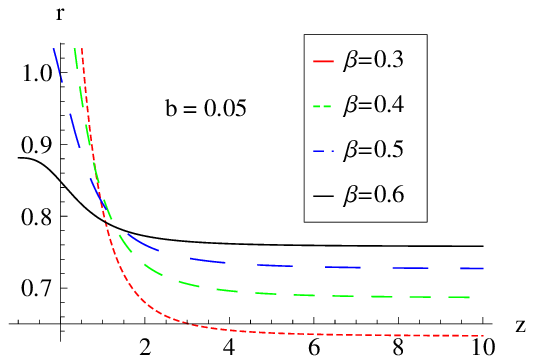}
\includegraphics[width=2.1 in]{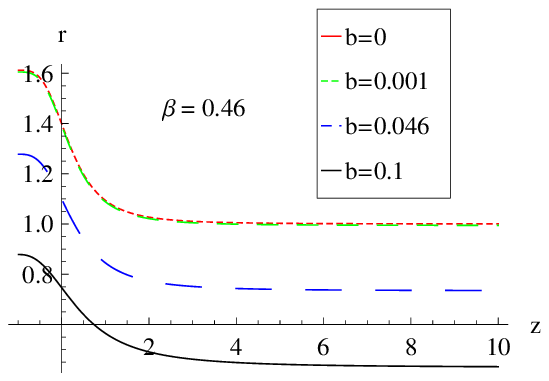}
\includegraphics[width=2.1 in]{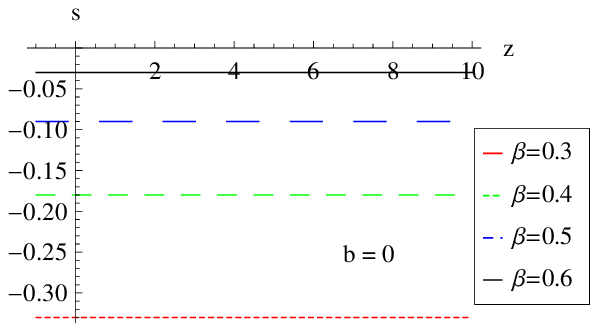}
\includegraphics[width=2.1 in]{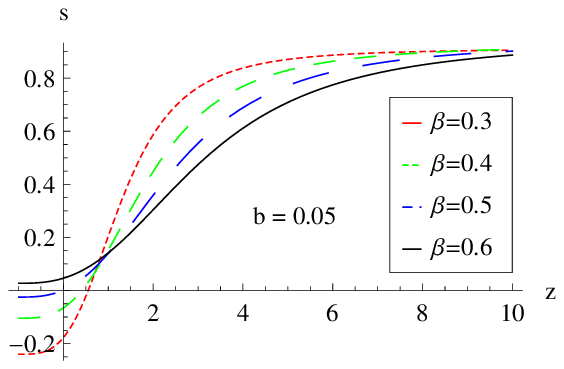}
\includegraphics[width=2.1 in]{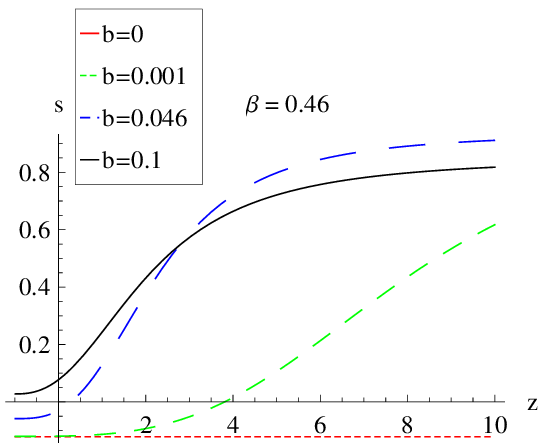}
\caption{(Color online.) The evolutions of the statefinder
parameters $r$ and $s$ with respect to the redshift $z$ in the ERDE
model without and with interaction.}\label{rs}
\end{figure*}

\begin{figure*}
\centering
\includegraphics[width=2.1 in]{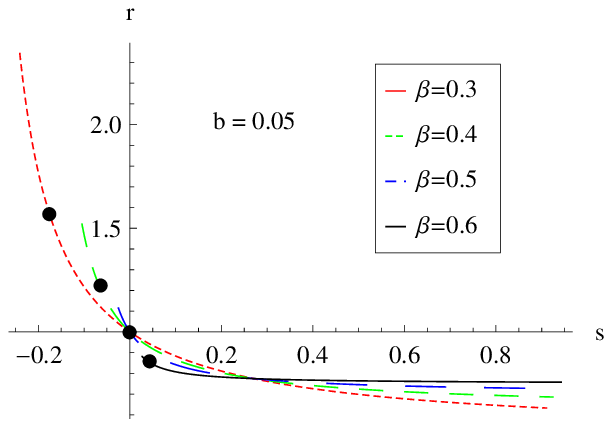}
\includegraphics[width=2.1 in]{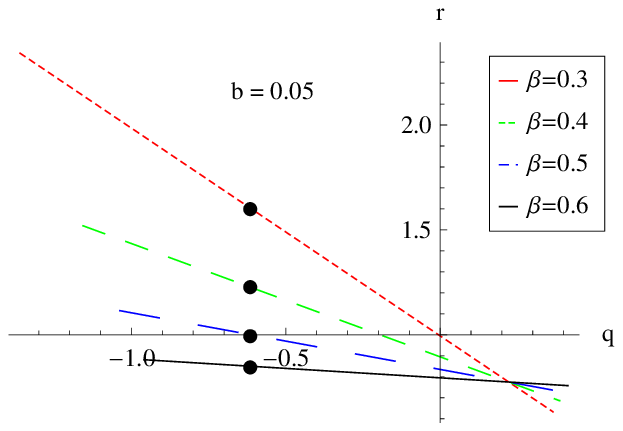}
\includegraphics[width=2.1 in]{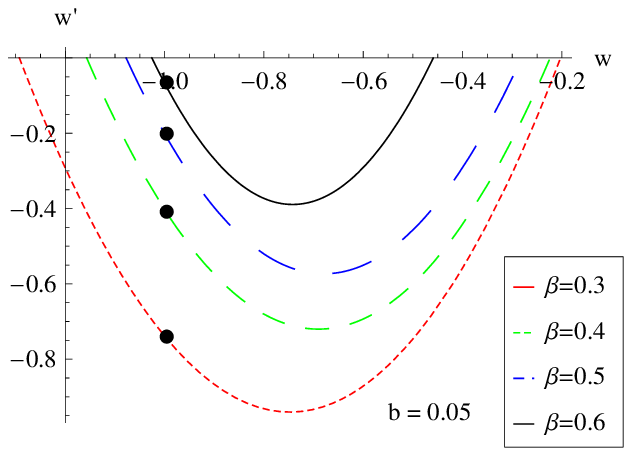}
\caption{(Color online.) The curves of $r(s)$, $r(q)$ and $w'(w)$ in
the ERDE model with interaction respectively in the $r-s$, $r-q$ and
$w'-w$ planes for variable $\beta$ with $b=0.05$. The dots denote
today's values of these parameters and $q_0=-0.595$ and $w_0=-1$ for
all the cases.}\label{beta}
\end{figure*}

Secondly, let us see the effect of the interaction on the ERDE model
from Fig.~\ref{bf}, where the best-fit values for parameters are
taken~\cite{Gao:2009prd043511,Feng:2008plb231,Yu:2010plb263,Zhang:2009un,Suwa:2010prd023519,Zhang:2010im,Fu:2011ab}.
In the $r-s$ plane, the interaction makes curves of $r(s)$ different
essentially. The curves for interacting cases can reach or tend to
the LCDM point $\{1,0\}$, even for very small $b$, and that both $r$
and $s$ are time dependent. Also the trajectories can lie in the
region $r<1$ and $s>0$, which are forbidden in the non-interacting
case (see the $r-s$ plane in Fig.~\ref{nin}). For the reason why
this phenomenon appears, we can see it from the evolutions of $r$
and $s$ with respect to the redshift $z$ in the ERDE model without
and with interaction in Fig.~\ref{rs}. Clearly the interaction can
make $r$ cross $r=1$ and $s$ cross $s=0$, so curves of $r(s)$ can
reach or tend to the LCDM point $\{1,0\}$, namely, the ranges of $r$
and $s$ are enlarged.

Finally, we plot curves of $r(s)$, $r(q)$ and $w'(w)$ in the general
HDE model with $b=0.05$ respectively in the $r-s$, $r-q$ and $w'-w$
planes in Fig.~\ref{beta}. We find that for the case of $\beta=0.5$
and $b=0.05$, the statefinder pair $\{r,s\}$ evolves just at the
LCDM point $\{1,0\}$ at present, as also can be recognized in
Fig.~\ref{rs}, i.e., when $z\sim 0$, $r_0\sim 1$ and $s_0\sim 0$.
Furthermore, we find that the smaller $\beta$ or $b$ is, the earlier
the interacting model reaches the LCDM point $\{1,0\}$ in the $r-s$
plane. So for the ERDE model, the introduction of interaction makes
its behavior more like other HDE models put forth before. That is to
say, the ERDE model with interaction is more favored.

\section{Conclusion}\label{sec:concl}
In this paper, we apply the statefinder diagnostic to the ERDE model
without and with interaction. In the non-interacting case, $r$ is
dependent of time while $s$ is a constant, like the QCDM models. But
the difference between them is that $w$ is time dependent in the
former while $w$ a constant in the latter. The non-interacting model
does not reach the LCDM point $\{1,0\}$ in the $r-s$ plane.
Especially for the case of $\alpha=2\beta$, it reduces to Ricci dark
energy and $\beta=0.5$ plays an important role in its
evolution~\cite{Feng:2008plb231}. In the interacting case, the
interaction changes in essence the behavior of the ERDE model,
because it makes $s$ no longer a constant and enlarges the ranges of
$r$ and $s$, then the LCDM point can be reached or tended to. This
is similar to other HDE models, so we can conclude that the ERDE
model with interaction is more favored. We hope that the future
high-precision observations can offer more and more accurate data to
determine these parameters precisely and consequently shed light on
the essence of dark energy.

\begin{acknowledgments}
This work was supported by the National Natural Science Foundation
of China under Grant Nos.~10975032, 11047112 and 11175042, and by
the National Ministry of Education of China under Grant
Nos.~N100505001 and N110405011.
\end{acknowledgments}



\begin{thebibliography}{*}
\bibitem{Riess:1998SN} A. G. Riess et al., Supernova Search Team Collaboration, Astron. J. {\bf 116} (1998) 1009.
\bibitem{Perlmutter:1999SN} S. Perlmutter et al., Supernova Cosmology Project Collaboration, Astrophys. J. {\bf 517} (1999) 565.
\bibitem{Spergel:2003WMAP} D. N. Spergel et al., WMAP Collaboration, Astrophys. J. Suppl. {\bf 148} (2003) 175.
\bibitem{Spergel:2007WMAP} D. N. Spergel et al., WMAP Collaboration, Astrophys. J. Suppl. {\bf 170} (2007) 377.
\bibitem{Adelman:2008SDSS} J. K. Adelman-McCarthy et al., SDSS Collaboration, Astrophys. J. Suppl. {\bf 175} (2008) 297.
\bibitem{Knop:2003SN} R. A. Knop et al., Supernova Cosmology Project Collaboration, Astrophys. J. {\bf 598} (2003) 102.
\bibitem{Komatsu:2009WMAP} E. Komatsu et al., WMAP Collaboration, Astrophys. J. Suppl. {\bf 180} (2009) 330.
\bibitem{Sahni:2000ijmpd373} V. Sahni and A. A. Starobinsky, Int. J. Mod. Phys. D {\bf 9} (2000) 373.
\bibitem{Peebles:2003rmp559} P. J. E. Peebles and B. Ratra, Rev. Mod. Phys. {\bf 75} (2003) 559.
\bibitem{Copeland:2006ijmpd1753} E. J. Copeland, M. Sami, and S. Tsujikawa, Int. J. Mod. Phys. D {\bf 15} (2006) 1753.
\bibitem{Li:2011sd} M. Li, X. -D. Li, S. Wang, and Y. Wang, Commun. Theor. Phys. {\bf 56} (2011) 525.
\bibitem{Sotiriou:2008rmp} T. P. Sotiriou and V. Faraoni, Rev. Mod. Phys. {\bf 82} (2010) 451.
\bibitem{Weinberg:1989rmp1} S. Weinberg, Rev. Mod. Phys. {\bf 61} (1989) 1.
\bibitem{Zlatev:1999prl896} I. Zlatev, L. Wang, and P. J. Steinhardt, Phys. Rev. Lett. {\bf 82} (1999) 896.
\bibitem{Caldwell:2002plb23} R. R. Caldwell, Phys. Lett. B {\bf 545} (2002) 23.
\bibitem{Feng:2005plb35} B. Feng, X. Wang, and X. Zhang, Phys. Lett. B {\bf 607} (2005) 35.
\bibitem{Padmanabhan:2002prd021301} T. Padmanabhan, Phys. Rev. D {\bf 66} (2002) 021301.
\bibitem{Piazza:2004jcap07004} F. Piazza and S. Tsujikawa, J. Cosmol. Astropart. Phys. {\bf 07} (2004) 004.
\bibitem{Hooft:9310026} G. 't Hooft, arXiv:gr-qc/9310026.
\bibitem{Cohen:1999prl4971} A. Cohen, D. Kaplan, and A. Nelson, Phys. Rev. Lett. {\bf 82} (1999) 4971.
\bibitem{Hsu:2004plb13} S. D. H. Hsu, Phys. Lett. B {\bf 594} (2004) 13.
\bibitem{Li:2004plb1} M. Li, Phys. Lett. B {\bf 603} (2004) 1.
\bibitem{Cai:2007plb228} R. G. Cai, Phys. Lett. B {\bf 657} (2007) 228.
\bibitem{Wei:2007ty} H. Wei and R. -G. Cai, Phys. Lett. B {\bf 660} (2008) 113.
\bibitem{Gao:2009prd043511} C. Gao, F. Wu, X. Chen, and Y. -G. Shen, Phys. Rev. D {\bf 79} (2009) 043511.
\bibitem{Granda:2008plb275} L. N. Granda and A. Oliveros, Phys. Lett. B {\bf 669} (2008) 275.
\bibitem{Sahni:2003jetpl201} V. Sahni, T. D. Saini, A. A. Starobinsky, and U. Alam, JETP Lett. {\bf 77} (2003) 201.
\bibitem{Alam:2003mnras1057} U. Alam, V. Sahni, T. D. Saini, and A. A. Starobinsky, Mon. Not. Roy. Astron. Soc. {\bf 344} (2003) 1057.
\bibitem{Zimdahl:2004grg1483} W. Zimdahl and D. Pav\'{o}n, Gen. Rel. Grav. {\bf 36} (2004) 1483.
\bibitem{Zhang:2005plb1} X. Zhang, Phys. Lett. B {\bf 611} (2005) 1.
\bibitem{Zhang:2005kj} X. Zhang, Commun. Theor. Phys. {\bf 44} (2005) 762.
\bibitem{Wu:2005ijmpd1873} P. Wu and H. Yu, Int. J. Mod. Phys. D {\bf 14} (2005) 1873.
\bibitem{Zhang:2005xk} X. Zhang, Commun. Theor. Phys. {\bf 44} (2005) 573.
\bibitem{Zhang:2004gc} X. Zhang, F. Q. Wu, and J. F. Zhang, J. Cosmol. Astropart. Phys. {\bf 01} (2006) 003.
\bibitem{Wei:2007plb1} H. Wei and R. -G. Cai, Phys. Lett. B {\bf 655} (2007) 1.
\bibitem{Chang:2007jcap01016} B. Chang, H. Liu, L. Xu, C. Zhang, and Y. Ping, J. Cosmol. Astropart. Phys. {\bf 01} (2007) 016.
\bibitem{Chang:2008mpla269} B. Chang, H. Liu, L. Xu, and C. Zhang, Mod. Phys. Lett. A {\bf 23} (2008) 269.
\bibitem{Shao:2008mpla65} Y. Shao and Y. Gui, Mod. Phys. Lett. A {\bf 23} (2008) 65.
\bibitem{Huang:2008ass175} Z. G. Huang, X. M. Song, H. Q. Lu, and W. Fang, Astrophys. Space Sci. {\bf 315} (2008) 175.
\bibitem{Panotopoulos:2008npb66} G. Panotopoulos, Nucl. Phys. B {\bf 796} (2008) 66.
\bibitem{Shojai:2009el30002} A. Shojai and F. Shojai, Europhys. Lett. {\bf 88} (2009) 30002.
\bibitem{Tong:2009prd023503} M. L. Tong and Y. Zhang, Phys. Rev. D {\bf 80} (2009) 023503.
\bibitem{Zhang:2010ijmpd21} L. Zhang, J. Cui, J. Zhang, and X. Zhang, Int. J. Mod. Phys. D {\bf 19} (2010) 21.
\bibitem{Zhang:2005ijmpd1597} X. Zhang, Int. J. Mod. Phys. D {\bf 14} (2005) 1597.
\bibitem{Setare:2007jcap03007} M. R. Setare, J. Zhang, and X. Zhang, J. Cosmol. Astropart. Phys. {\bf 03} (2007) 007.
\bibitem{Zhang:2008plb26} J. Zhang, X. Zhang, and H. Liu, Phys. Lett. B {\bf 659} (2008) 26.
\bibitem{Feng:2008plb231} C. -J. Feng, Phys. Lett. B {\bf 670} (2008) 231.
\bibitem{Yu:2010plb263} F. Yu, J. Zhang, J. Lu, W. Wang, and Y. Gui, Phys. Lett. B {\bf 688} (2010) 263.
\bibitem{Li:2009jcap12014} M. Li, X. -D. Li, S. Wang, Y. Wang, and X. Zhang, J. Cosmol. Astropart. Phys. {\bf 12} (2009) 014.
\bibitem{Komatsu:10014538} E. Komatsu et al., Astrophys. J. Suppl. {\bf 192} (2011) 18.
\bibitem{Zhang:2009un} X. Zhang, Phys. Rev. D {\bf 79} (2009) 103509.
\bibitem{Suwa:2010prd023519} M. Suwa and T. Nihei, Phys. Rev. D {\bf 81} (2010) 023519.
\bibitem{Zhang:2010im} J. Zhang, L. Zhang, and X. Zhang, Phys. Lett. B {\bf 691} (2010) 11.
\bibitem{Fu:2011ab} T. -F. Fu, J. -F. Zhang, J. -Q. Chen, and X. Zhang, Eur. Phys. J. C {\bf 72} (2012) 1932.
\end{thebibliography}
\end{document}